\documentclass[12pt]{iopart}

\usepackage{graphicx}
\usepackage{float}
\usepackage{caption}
\usepackage{hyperref}
\usepackage{comment}

\begin{document}

\title[]{Preliminary mapping of ionospheric total electron content (TEC) over Ecuador using global positioning system (GPS) data}

\author{E. D. López$^{1,2}$, B. A. Ubillús$^{1}$, A. A. Meza$^{1}$}

\address{$^{1}$ Escuela Politécnica Nacional, Facultad de Ciencias, Departamento de Física, Quito-Ecuador.}
\address{$^{2}$Escuela Politécnica Nacional, Observatorio Astronómico de Quito, Quito-Ecuador}
\ead{ericsson.lopez@epn.edu.ec}
\vspace{3pt}

\begin{abstract}
The ionosphere affects radio signals by altering their speed, direction, and trajectory, causing a temporary delay known as ionospheric delay, which is directly related to the total electron content (TEC). Although research in other equatorial locations has explored TEC implications, qualitative research is required to predict its behavior in the ionosphere. This study aims to depict TEC intensity evolution through color maps using data from 14 GPS receivers placed across Ecuador. For this purpose, pseudorange observables collected from the stations were used to present a calculation method for the TEC and its evolution during January 2022. The results revealed an oscillatory behavior in the evolution of the TEC, with intensity peaks that, on some occasions, approach and even exceed 100 $TECU$ (TEC units), while its local minimums never reach zero values.
\end{abstract}

%
%
%
%
%

\section{Introduction}

GPS accuracy is based on the meticulous synchronization of all signal components by means of atomic clocks. These clocks have long-term frequency stability of $10^{-13}$ and $10^{-14}$ over the course of a day. GPS satellites have these high-precision clocks, also known as frequency standards, which generate a fundamental frequency of 10.23 MHz. From this frequency, two signals are obtained, the carrier waves $f_1$ and $f_2$, which are generated by multiplying the fundamental frequency by the values 154 and 120, respectively $f_1 = 1575.42 \mbox{ MHz}$, $f_2=1227.6 \mbox{ MHz.}$ \cite{Ubillus, Hofmann} \\

When radio waves, such as those emitted by GPS satellites, pass through the ionosphere, they experience two notable effects: deflection in their path and delay in the arrival of the signal.
These phenomena are caused by the presence of free electrons in the ionosphere, which produce the refraction effect governed by Snell's law (see \cite{Hargreaves}, p. 145). However, the dynamics of waves within the ionosphere cannot be solely explained by this simple expression. To gain a comprehensive understanding of how radio waves behave when passing through this layer, it is essential to consider that it constitutes a plasma composed of different regions \cite{Hargreaves, Komjathy}, with a wide variety of unevenly distributed irregularities.

Edward Appleton developed the magneto-ionospheric theory to describe the refractive index in the ionosphere. In his research, he showed that when a plane-polarized wave passes through a magnetized plasma, it splits into two circularly polarized waves rotating in opposite directions. Subsequently, Douglas Hartree suggested the application of Lorentz polarization in the theorem, which led to the Appleton-Hartree formulation for calculating the complex refractive index \cite{Markovic}. The formulation and its development are presented below.


The equatorial region of the ionosphere is characterized by having the highest values of electron density. In this area, the magnetic field extends horizontally, which prevents the flow of current in the vertical direction that would normally occur. This generates the appearance of charges at the upper and lower limits, creating an electric field that increases the current in the horizontal plane. It has been established that, in this special circumstance, the conductivity along the magnetic field and in a horizontal direction is known as \textit{Cowling conductivity} \cite{Glassmeier}, which has a value comparable to the direct conductivity, resulting at an abnormally high current along the magnetic equator. This intense current is known as an equatorial ``electrojet'' \cite{Ericson}.


The total electron content (TEC) is a descriptive parameter of the ionosphere that represents the amount of electrons along the direct path of the signal between the receiver and the satellite \cite{Ericson, Toapanta}. By definition, the TEC is the integral of the electron density $N(s)$ along a path $ds$ between points A and B: $TEC = \int^{B}_{A} N(s) ds.$
TEC is generally measured in TECU units (1 TECU = $10^{16}$ $e$/$m^2$) and is estimated using GPS pseudorange and carrier phase data \cite{ Japan, Scharroo}. \\

To facilitate the comparison of the electron concentration in trajectories with different elevation angles, it is necessary to convert the TEC values into their vertical equivalent ($TEC_v$) \cite{Mannucci, Markovic, Ericson}. This parameter represents the total number of electrons in a column perpendicular to the ground.

Using a simple mapping function, slant TEC values are transformed to vertical TEC values, which are associated with a specific latitude and longitude corresponding to the ionospheric pierce point \cite{Hargreaves}. This transformation is performed under the assumption that the ionosphere can be approximated as a thin layer compressed at a maximum altitude of 350 $km$. In general, the conversion can be performed using the  expression
$ TEC_{v} = TEC_{s} ~cos(z')$. On the other hand, $z'$ can be found from  $sen(z') = \frac{R_E}{R_E + h_m} sen(z)$, where $z$ and $z'$ ($z' = 90^{\circ}$-$z$) are the zenith angles at the receiver and at the ionospheric drilling point respectively, $R_E$ is the radius Earth average and $h_m$ is the height of the maximum electron density. The usual values are $R_E=6371$ $km$ and $h_m=506.7$ $km$ \cite{Markovic}.

\section{TEC measurements from GPS signals} \label{segundo}

\subsection{Ionospheric refractive index}

We begin by considering some important points regarding the Appleton-Hartree equation. First of all, it is essential to emphasize that this formula is applied in an electrically neutral environment, without the presence of space charges, and with an equilibrium between electrons and cations. Furthermore, a constant magnetic field is considered, and the effect of the cations on the wave is assumed to be negligible \cite{Markovic, Ericson}. \\

To find the ionospheric refractive index using the Appleton-Hartree equation, we assume a plane electromagnetic wave traveling along the $x$ axis of the orthogonal coordinate system in the presence of a uniform external magnetic field that makes an angle $\theta$ with the direction of wave propagation. In this context, the complex refractive index $n$ of the Appleton-Hartree theory is expressed as follows \cite{Helliwell}:

\vspace{0.5em}

\begin{equation} \label{eb}
    n^2=1-\frac{X}{\left( 1-iZ\right)-\left[ \frac{Y^2_T}{2\left( 1-X-iZ\right)}\right]\pm \left[ \frac{Y^4_T}{4\left( 1-X-iZ\right)^2} + Y^2_L\right]^{\frac{1}{2}}},
\end{equation}

\vspace{0.5em}

\noindent with $
     X=\frac{\omega^2_N}{\omega^2} = \frac{f^2_N}{f^2}, 
     Y=\frac{\omega_H}{\omega}=\frac{f_H}{f}, 
     Y_L=\frac{\omega_L}{\omega} = \frac{f_L}{f},  Y_T=\frac{\omega_T}{\omega}=\frac{f_T}{f}, 
     Z=\frac{\omega_c}{\omega}=\frac{f_c}{f}$.

\vspace{0.5em}

\noindent where, $\omega$ is the angular frequency of the carrier wave,
     $f$ is the frequency of the carrier wave with $f=\frac{\omega}{2\pi}$. Analogously, to the rest of the angular frequencies,
     $\omega_N$ is the angular frequency of the plasma, which is calculated by the formula $\omega^2_N= \frac{Ne^2}{\varepsilon_0 m_e}$, with electron density $N$, electronic charge $e$, vacuum dielectric permittivity $\varepsilon_0$ and electronic mass $m_e$,
     $\omega_H$ is the cyclotron angular frequency of the free electrons, which is calculated with the formula $\omega_H = \frac{B_0|e|}{m_e}$, with magnetic induction $B_0$,
    $\omega_T$ is the transverse component of $\omega_H$, which is calculated by the formula $\omega_T=\frac{B_0|e|}{m_e}sin(\theta)$,
     $\omega_L$ is the longitudinal component of $\omega_H$, which is calculated with the formula $\omega_L=\frac{B_0|e|}{m_e}cos(\theta)$ and
    $\omega_c$ is the angular frequency of collisions between electrons and heavy particles.\\

\noindent For the case in which the collisions are negligible ($Z\approx 0$), and the influence of the magnetic field is neglected, the equation \ref{eb} is reduced to $  n^2\approx 1-\frac{2X(1-X)}{2(1-X)-Y^2_T\pm \left[ Y^4_T + 4(1-X)^2 Y^2_L \right]^{\frac{1}{2}}}$ $(*)$.\\

\vspace{0.5em}

According to the magneto-optical theory, when a plane-polarized electromagnetic wave passes through a medium with a magnetic field, it is split into two different waves \cite{Hargreaves, Markovic}. The first wave, known as an \textit{ordinary wave}, behaves similarly to a wave that propagates in the absence of a magnetic field. This is represented in the equation (*) with a ``+'' sign. On the other hand, the second wave called \textit{extraordinary wave}, is characterized by having a ``-'' sign in the equation. \\

Expanding equation (*) by the Taylor series, and if the influence of the magnetic field is neglected ($\theta \approx 0$), only the first two terms of the expansion can be used. In this case, the refractive index of the ionospheric plasma used to calculate the TEC is given simply by:

\begin{equation}\label{e3}
    n=1-\frac{1}{2}X = 1-\frac{1}{2}\frac{f^2_N}{f^2},
\end{equation}

\vspace{0.5em}

\noindent where $f^2_N$ as a function of $N$ is $f^2_N=80.6N$ $Hz^2$. Therefore, for the ionosphere, the phase refractive index appropriate for carrier phase observations can be written as follows $n_p = 1-40.3 \frac{N}{f^2},$  while the group refractive index, suitable for pseudorange observations, can be determined by $ n_g=1+40.3 \frac{N}{f^2}$.
The delay in the propagation of electromagnetic waves, caused by the refractive index (\ref{e3}), constitutes the main source of error in the GPS positioning system. This phenomenon causes a delay in the arrival of the waves, which generates inaccuracies in determining the exact location of the user \cite{Webster}.

\subsection{Ionospheric delay}

The delay of a signal passing through the ionosphere is defined as the integral of the refractive index $n$ along the path $ds$ extending from the satellite ($Sat$) to the receiver ($Rec$):
$S=\int^{Rec}_{Sat} n ds$.
Replacing here the index of refraction $n_p$ gives

\vspace{0.5em}

\begin{equation}\label{e7}
     S=\rho-40.3 \frac{1}{f^2}\int_{Sat}^{Rec} N ds = \rho - 40.3 \frac{TEC}{f^2},
\end{equation}

\vspace{0.5em}

\noindent where $\rho$ is the distance between the satellite and the receiver (without delays). In the last equality, the definition of the TEC  was used to replace the path integral. On the other hand, the modulated signal can be expressed equivalently as $S=\rho + 40.3 \frac{TEC}{f^2}$.\\

From the two previous expresions for $S$, it can be inferred that when passing through the ionosphere, the phase of the carrier wave experiences an advance (\ref{e7}), because the distance $S$ is shorter than the real distance $\rho $, while the modulated signal experiences a delay because the distance $S$ is greater than the actual distance $\rho$. The term that is subtracted and added in the respective expressions corresponds to the difference between the real distance $\rho$ and the distance $S$ and represents the error caused by the propagation of the signal through the ionosphere \cite{ Webster, Garner, Ismail}. This error is known as \textit{ionospheric delay} ($
     d_{ion} = 40.3 ~\frac{TEC}{f^2}$).

\vspace{0.5em}

By knowing the frequencies of the GPS signal, it can be established that the delay of the ionospheric signal is exclusively dependent on the TEC. Therefore, knowing the TEC and its characteristics, the ionosphere can be modeled for scientific purposes; for example, the prediction of solar storms based on the development and changes in the number of electrons, as well as the determination of the propagation error in radio waves \cite{Nishioka}.

\subsection{Pseudorange TEC}

For a specific signal of frequency $f_i$ with $i=1,2$, the pseudorange $P$ can be expressed using the reception time $t_r$ (measured by the receiver clock) and the transmission time $t_s$ (measured by the satellite clock), as $ P_i=c \left( t_r-t_s \right), \quad i={1,2}$. This equation can be rewritten for the frequencies $f_1$ and $f_2$ (previously presented) as

\vspace{0.5em}

\begin{equation}\label{e10}
     P_1=\rho + c\left[ dt(t_r)-dT(t_s) \right] + d_{ion_{1}} + d_{trop} + \epsilon_{p_1},
\end{equation}

\vspace{0.5em}

\begin{equation}\label{e11}
     P_2=\rho + c\left[ dt(t_r)-dT(t_s) \right] + d_{ion_{2}} + d_{trop} + \epsilon_{p_2},
\end{equation}

\vspace{0.5em}

\noindent where: $\rho$ is the range (without delays) between the satellite and the receiver,
   $c$ is the speed of light,
     $dt(t_r)$ is the receiver clock offset,
    $dT(t_s)$ is the satellite clock offset,
     $d_{ion}$ is the ionospheric delay,
      $d_{trop}$ is the tropospheric delay, and
      $\epsilon_p$ represents receiver and satellite instrumental delays, measurement noise including satellite orbital errors, and thermal noise. Subtracting the equation (\ref{e11}) from (\ref{e10}) we obtain

\vspace{0.5em}

\begin{equation}
     P_1-P_2 = d_{ion_{1}}-d_{ion_{2}} + \left( \epsilon _{p_1} - \epsilon _{p_2} \right),
\end{equation}

\vspace{0.5em}

\noindent from which the term $\left( \epsilon _{p_1} - \epsilon _{p_2} \right)$ can be neglected since its contribution to the TEC is insignificant. So, $P_1-P_2= d_{ion_{1}}-d_{ion_{2}}$. Substituting the expressions for $d_{ion_{1}}$ and $d_{ion_{2}}$ we arrive at

\vspace{0.5em}

\begin{equation*}
     P_1-P_2= 40.3\frac{TEC_p}{f^2_1}-40.3\frac{TEC_p}{f^2_2}.
\end{equation*}

\vspace{0.5em}

\noindent Consequently, the TEC of the pseudorange is calculated by:

\vspace{0.5em}

\begin{equation}\label{e12}
     TEC_p= \frac{1}{40.3}\left( \frac{f_1 f_2}{f_1-f_2} \right) \left( P_2-P_1 \right).
\end{equation}

\vspace{0.5em}

The TEC between the satellite and the user, which varies depending on the elevation angle of the satellite, is known as \textit{oblique} or \textit{slant TEC} ($TEC_s$) \cite{Markovic, Jakowsi}. This value represents the total electron density in the ionosphere along the beam path from the satellite to the receiver.

\subsection{GPS data}

The Geodetic Continuous Monitoring Network of Ecuador (REGME) generates 24-hour files daily in its primary raw data format, using a time interval of one second and following the RINEX 2.11 standard. Each station records and stores the pseudo-range and carrier phase in files with recording intervals of 30 seconds, these values being the main ones for the calculation and analysis of the respective TEC. To collect the aforementioned data, we proceed to download the files from the official Military Geographic Institute of Ecuador (IGM) page.  The 14 GPS receivers used in this work are shown in the following figure $\ref{figure:6}$:

\begin{figure}[H]
\centering
\includegraphics[scale=0.66]{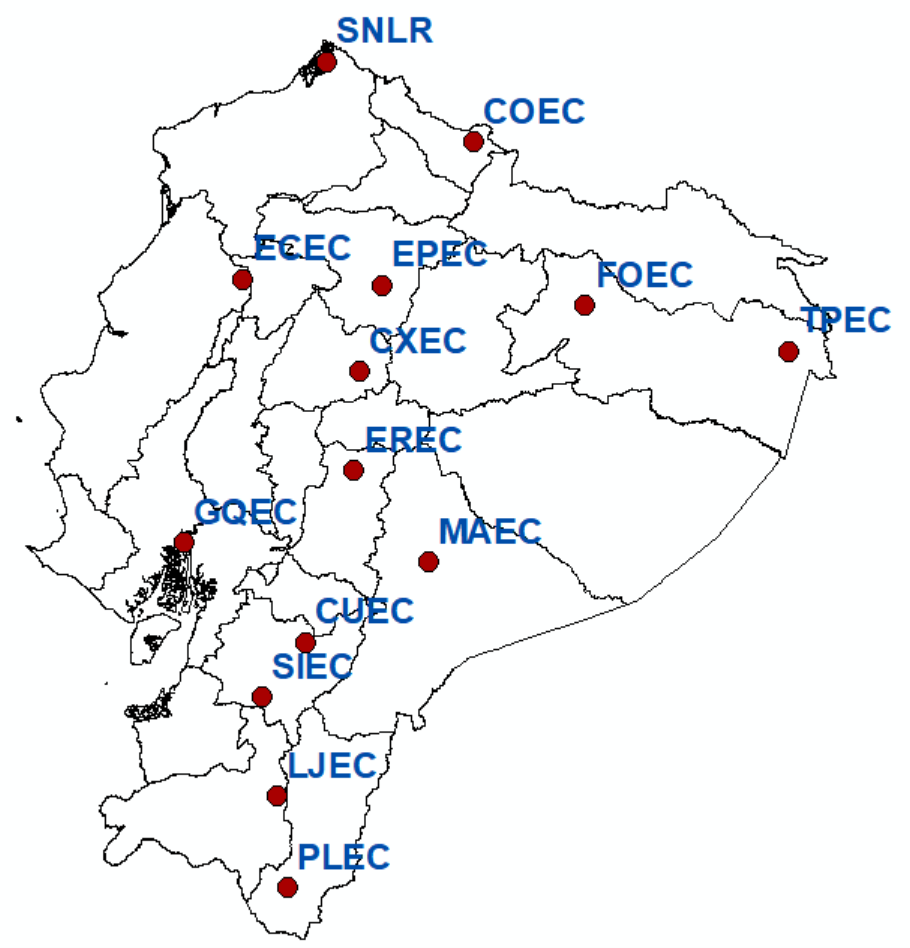}
\caption{Location of GPS continuous monitoring stations over Ecuador}
\label{figure:6}
\end{figure}

Using a Fortran program the format conversion was carried out, transforming the RINEX 2.11 files to compact RINEX \cite{Rideout}, to later calculate the TEC and save the results in a .txt file with two columns: the first shows the date and the second presents the corresponding TEC values. Subsequently, another C++ program was implemented to change the $TEC_s$ for the $TEC_v$.

\section{TEC mapping results and discussion}

\begin{figure}[H]
\centering
\includegraphics[scale=0.305]{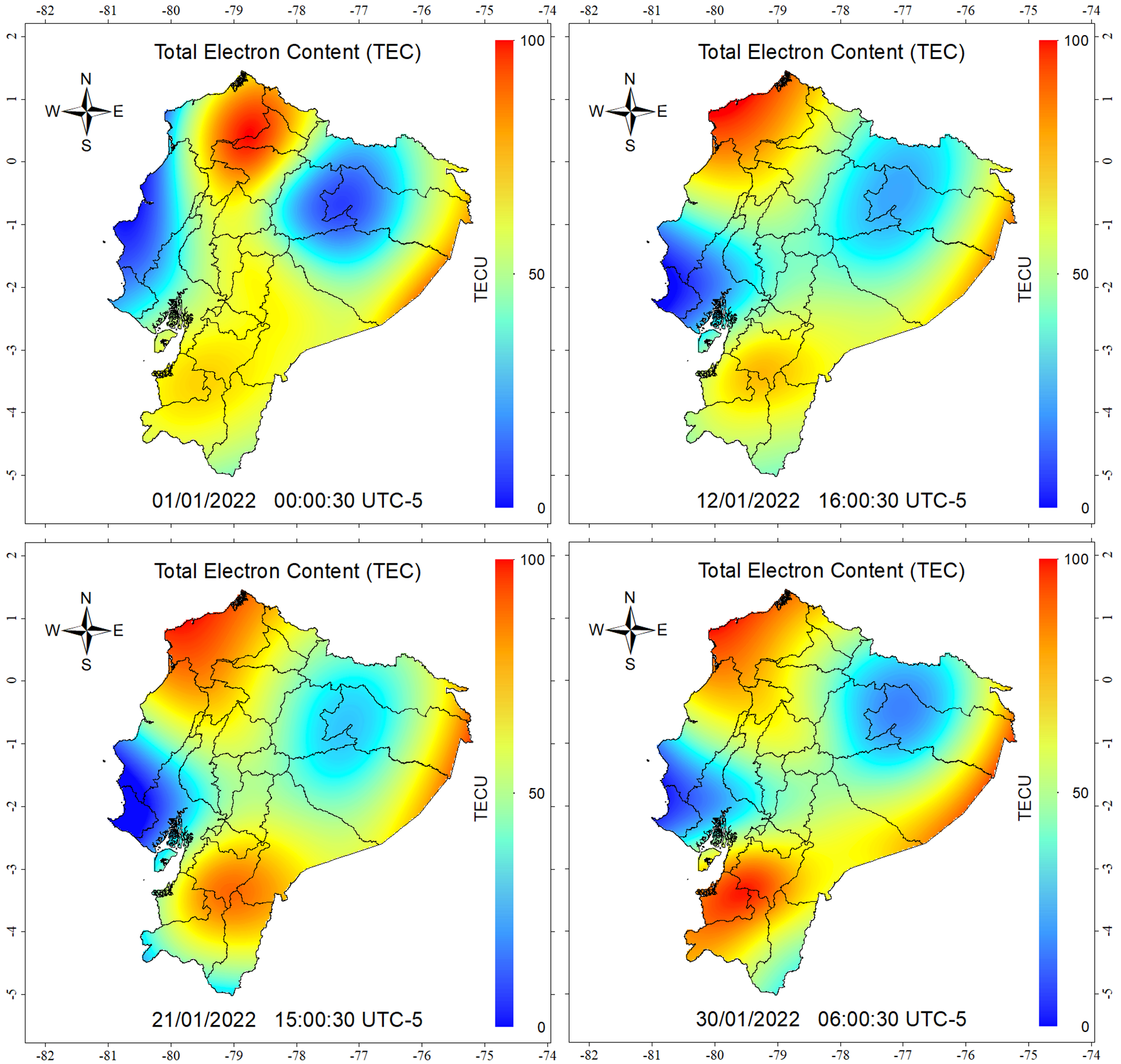}
\caption{TEC maps for four different dates. TEC Animation over Ecuador - January 2022: \url{https://youtu.be/o6S0mlPYCsM}}
\label{fig:mapa}
\end{figure}

To create the TEC map over Ecuador, the .txt files with TEC data were exported to the software ArcGIS. After completing this step, the Spline interpolation tool is used by specifying the exported file and the column of data to interpolate to; in this case, the ``TEC'' column is considered. As a result, a map of Ecuador is obtained showing the TEC intensity by means of a color palette, as shown in Figure \ref{fig:mapa}. This map corresponds to a specific time instant of a certain day, which is why this procedure must be repeated for each hour of the day. After obtaining all the maps, we proceed to create an animation showing the vertical TEC intensity in the whole region during January 2022. \\

The spatial coverage of the maps ranges from -5$^\circ$ to 2$^\circ$ N latitude and from -82$^\circ$ to -74$^\circ$ E longitude. Data were recorded with a temporal resolution of 1-hour intervals. \\

Figure \ref{fig:mapa} shows that the TEC reaches values of 100 $TECU$, which is considered high compared to other geographical locations. This high electron concentration is mainly attributed to the electric field storms in the equatorial region, which generate the electrojet effect, resulting in a higher electron density in this area. \\

When analyzing the evolution of the maps in Figure \ref{fig:mapa}, a similar behavior between image transitions is evidenced, except for slight variations in concentration in specific areas of the country. In the Sierra region, especially in the southern zone, a remarkable intensity of electrons is observed, while in the central zone of the region an intermediate density is recorded. On the other hand, in the Coast and East regions, a lower concentration is observed compared to the Sierra, except in the province of Esmeraldas, where high amounts of electrons were obtained. This reveals a similar pattern in the variation of TEC along the entire surface, which implies that the evolution of this parameter, in each of the GPS stations, exhibits the same behavior, with its local extremes (either peaks or valleys) coinciding at the same time of the day. In addition, these maxima and minima present variations in their magnitude, indicating a greater and lesser intensity of the TEC in the previously mentioned regions. \\

On the other hand, on days 01 and 08, TEC behavior was different from the other days of January. One of the possible explanations is that transient atmospheric phenomena, such as solar flares, could have contributed to the incident solar radiation, affecting the conditions of the ionosphere. In addition, geomagnetic activity, which can vary due to factors such as the interaction between the solar wind and the Earth's magnetic field, could also have altered the TEC on those days. Another factor to consider is the interaction between local climatic characteristics and the ionosphere. Since this layer responds to atmospheric conditions, the meteorological phenomena could have contributed to the TEC variations present in the maps. All these events can induce changes in the electron density in the ionosphere, generating TEC maxima and minima in areas different from those observed on subsequent days.

It is important to mention that the maps were generated using an interpolation method, which implies that the blue areas in Figure \ref{fig:mapa} do not necessarily indicate a low TEC ($\approx$ $0 - 10$ TECU), but rather represent locations where the TEC is relatively lower compared to the orange and red areas, which reflect a higher electron density. \\

Below are time series graphs showing the evolution of the TEC over one day, one week and one month.

\subsection{Daily}

\begin{figure}[H].
\centering
\includegraphics[scale=0.57]{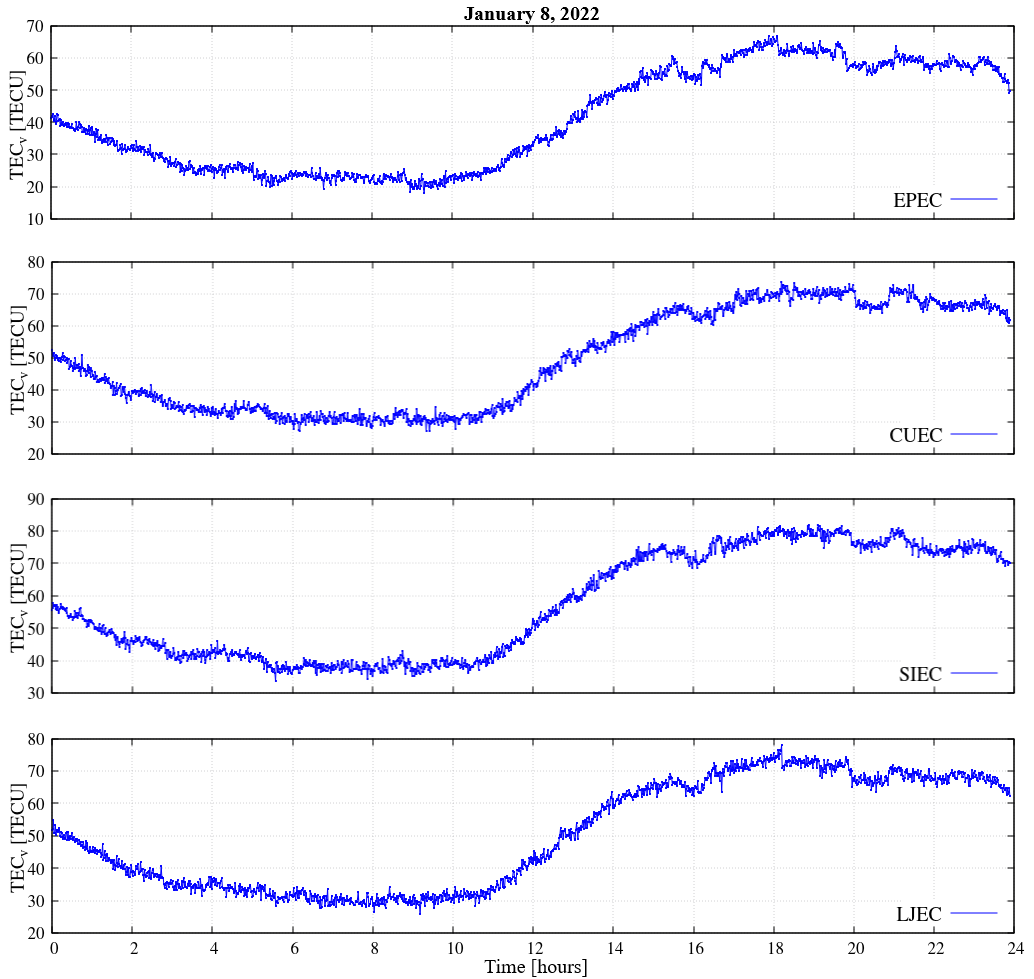}
\caption{Variation of TEC during one day at four different GPS receivers}
\label{figure:12}
\end{figure}

The diurnal variation of TEC can be seen in Figure \ref{figure:12}. A gradual decrease is observed at the beginning of the day (00:00 UTC-5) until reaching a minimum at 08:00 UTC-5, from which, an increase is presented until reaching a maximum in the curve around 18:00 UTC-5, followed again by a decrease. This behavior is mainly attributed to the influence of solar radiation during the day and its absence at night. This is specifically due to the fact that in the morning solar radiation ionizes the atmosphere, resulting in a (non-immediate) increase in TEC. In the evening hours the lack of solar radiation reduces atmospheric ionization, causing a decrease in electron density.

\subsection{Weekly}

\begin{figure}[H].
\centering
\includegraphics[scale=0.57]{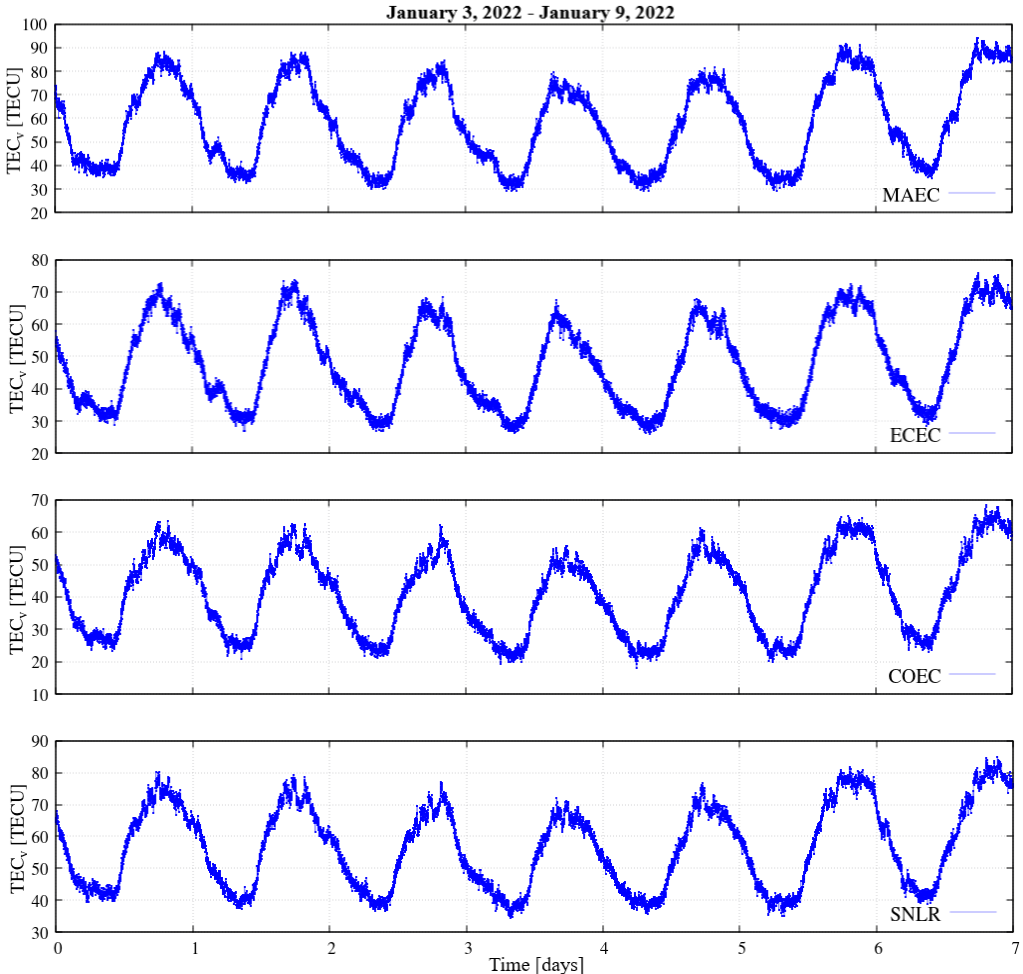}
\caption{Variation of TEC during one week in four different GPS receivers}
\label{figure:13}
\end{figure}

Figure \ref{figure:13} presents the weekly TEC evolution which is similar to the diurnal change, but on a longer time scale. In this graph, the TEC variation over time can be better observed, revealing an oscillatory behavior caused by solar activity.

\subsection{Monthly}

\begin{figure}[H].
\centering
\includegraphics[scale=0.57]{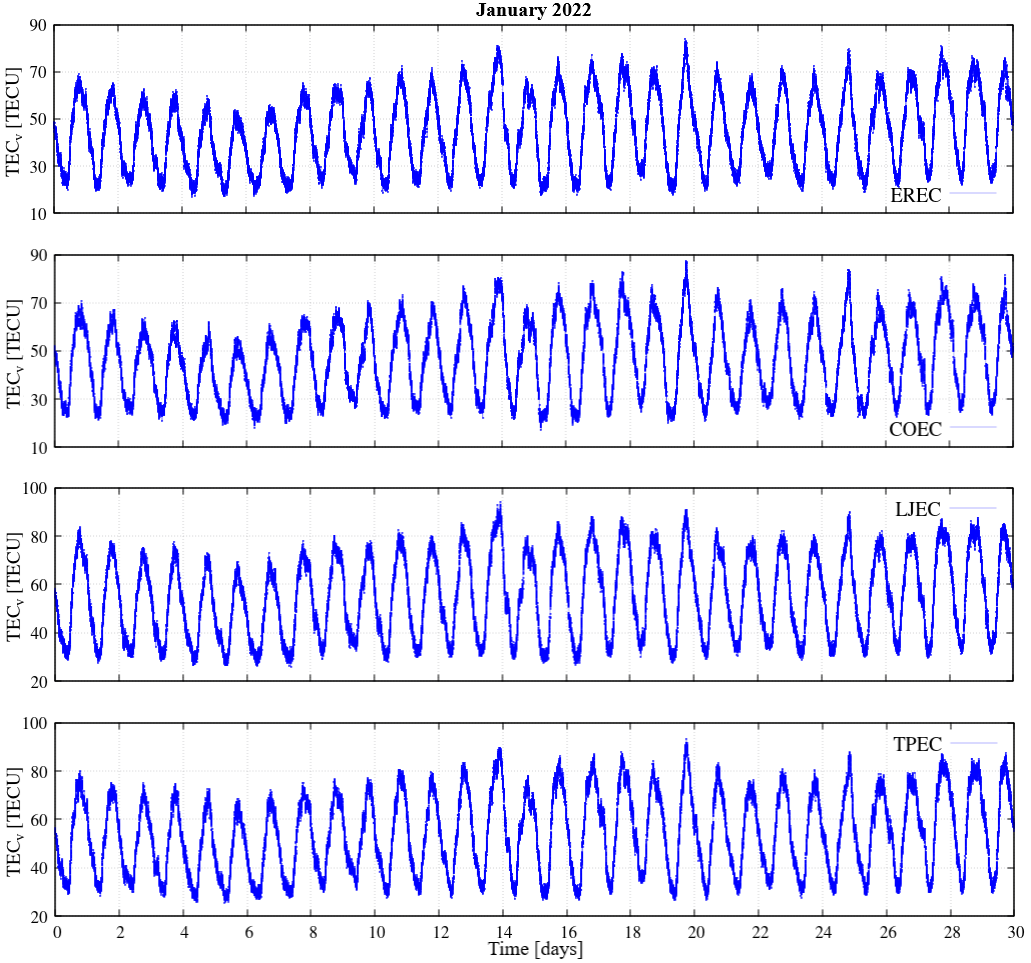}
\caption{Variation of TEC during one month in four different GPS stations}
\label{figure:14}
\end{figure}

For a time interval of one month (see Figure \ref{figure:14}), the TEC oscillations are similar to the weekly case presented in Figure \ref{figure:13} but more evident. It is interesting to note that the TEC never reaches zero values at any point due to the dynamic nature of the ionosphere; that is, as the TEC tends toward near-zero values, solar radiation begins to recharge the ionosphere at dawn thus preventing the values from reaching completely zero. In addition, anomalies characteristic of the $F2$ region also contribute to this phenomenon. Unlike other regions of the ionosphere that can fade during the night, the $F2$ region maintains significant electron levels until the next sunrise. 

\begin{figure}[H]
\centering
\includegraphics[scale=0.725]{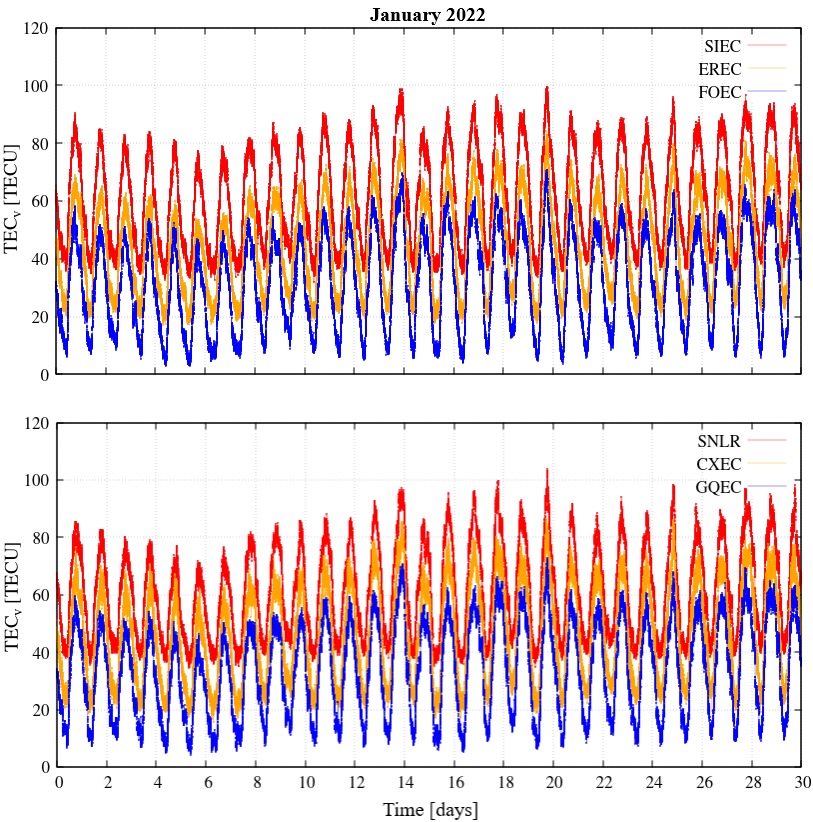}
\caption{TEC difference between various GPS stations}
\label{fig:15}
\end{figure}

Figure \ref{fig:15} reveals that the local extrema of each GPS station coincide at the same time of day; that is, they are not out of phase with each other. This indicates that the TEC, throughout the territory, varies similarly. On the other hand, a significant difference in TEC magnitude is observed in the SIEC (Azuay) - EREC (Chimborazo) - FOEC (Morona Santiago) and SNLR (Esmeraldas) - CXEC (Cotopaxi) - GQEC (Guayaquil) stations. This variation of approximately 10 TECU is attributed to the amount of incident solar radiation in these areas, showing that the provinces of Esmeraldas and Cuenca experience a higher incidence of electromagnetic radiation, while in Morona Santiago and Guayaquil there are times when sunlight is much dimmer. For this reason, most of the maps show high electron densities in the north and south of the country.

\section{Conclusions}

The maps allowed a detailed visualization of the spatial distribution of the TEC at one-hour intervals throughout the month of January. It was seen that the TEC never reaches zero values due to the constant interaction between solar radiation and the ionosphere, as well as the characteristic anomalies of the F2 region that maintain significant levels of electrons, even during the night. In addition, TEC values of up to 100 TECU were recorded at several equatorial GPS stations, verifying the fact that the equatorial region has the highest electron density levels. These values highlight the importance of monitoring and understanding TEC dynamics in the region. \\

On the other hand, the TEC maps and their time series confirmed that the electron density in the ionosphere is directly related to the incident solar radiation, leading to the occurrence of daily maxima and minima during its evolution. This behavior is represented on longer time scales such as weeks and months, where an oscillatory variation of the TEC is observed, with higher peaks and lower valleys in certain areas. These variations are also closely linked to solar activity, which contributes more to the electron density in specific locations such as Azuay (SIEC receiver) and Esmeraldas (SNLR receiver), increasing the electron concentration in those areas. \\

The results provide valuable information on the intensity of electrons over the entire surface of the equator, allowing their movement in the ionosphere to be tracked. By tracking their trajectory, the equatorial geomagnetic field conditions can be predicted, which has important implications for the correction of ionospheric refraction in communication systems and for estimating the signal delay between receivers and satellites, thus improving the accuracy of GPS positioning. \\

Visual representations of TEC are especially relevant in the prediction of ionospheric phenomena, such as solar storms and geomagnetic disturbances. These events originating from solar activity can generate significant changes in the ionosphere, affecting the propagation properties of GPS signals. By continuously monitoring the TEC through mapping, it is possible to detect these changes in the ionosphere, identify possible space weather events, and anticipate their effects on satellite communications and navigation systems. \\

In the field of signal propagation studies, TEC maps are useful for understanding how variations in electron density impact electromagnetic wave propagation, especially at frequencies from 3 KHz to 30 GHz. These changes can cause attenuation of radio signals as they pass through the ionosphere. The maps allow the identification of regions where waves may experience greater attenuation, making it possible to take measures to mitigate the adverse effects and improve the quality of communications. \\

\section*{References}

    \bibliographystyle{unsrt}
    \bibliography{referencias}

\end{document}